\documentclass[smallextended]{svjour3}
\usepackage{lmodern}
\usepackage[T1]{fontenc}
\usepackage[utf8]{inputenc}
\usepackage[colorlinks=true,allcolors=blue]{hyperref}
\usepackage{graphicx}
\usepackage{tensor}
\usepackage{mathrsfs}
\usepackage{enumitem}
\usepackage{latexsym}
\def\nobibtexrefs{} 

\newcommand{\be}{\begin{equation}}
\newcommand{\ee}{\end{equation}}
\newcommand{\bea}{\begin{eqnarray}}
\newcommand{\eea}{\end{eqnarray}}

\newcommand{\nnm}{\nonumber}

\newcommand{\qq}{\quad\quad}
\newcommand{\qqq}{\qquad\quad}
\newcommand{\ve}{\varepsilon}

\journalname{Gen.~Relativ.~Gravit.}

\begin{document}

\title{Properties of an affine transport equation and its holonomy}
\author{Justin Vines \and David A.\ Nichols}
\institute{J.~Vines \at Max-Planck-Institut f\"ur Gravitationsphysik,
Albert-Einstein-Institut Am M\"uhlenberg 1, 14476 Golm, Germany.
\email{justin.vines@aei.mpg.de} \and
D.~A.~Nichols \at Cornell Center for Astrophysics and Planetary 
Science, Cornell University, Ithaca, New York 14853, USA.
\email{david.nichols@cornell.edu}}
\date{Received: \today}

\maketitle

\begin{abstract}
An affine transport equation was used recently to study properties of
angular momentum and gravitational-wave memory effects in general relativity.
In this paper, we investigate local properties of this transport equation in 
greater detail.
Associated with this transport equation is a map between the tangent spaces 
at two points on a curve.
This map consists of a homogeneous (linear) part given by the parallel 
transport map along the curve plus an inhomogeneous part, which is related 
to the development of a curve in a manifold into an affine tangent space.
For closed curves, the affine transport equation defines a 
``generalized holonomy'' that takes the form of an affine map on the tangent
space.
We explore the local properties of this generalized holonomy by using 
covariant bitensor methods to compute the generalized holonomy around 
geodesic polygon loops.
We focus on triangles and ``parallelogramoids'' with sides formed from 
geodesic segments.
For small loops, we recover the well-known result for the leading-order 
linear holonomy ($\sim$ Riemann $\times$ area), and we derive the leading-order
inhomogeneous part of the generalized holonomy 
($\sim$ Riemann $\times$ area$^{3/2}$). 
Our bitensor methods let us naturally compute higher-order corrections to 
these leading results.
These corrections reveal the form of the finite-size effects that enter into 
the holonomy for larger loops; they could also provide quantitative errors on 
the leading-order results for finite loops.
\end{abstract}

\section{Introduction}

In general relativity, vectors and tensors are most often transported between 
tangent spaces at different spacetime points using the Levi-Civita connection
on the tangent bundle (the unique connection that is both metric-compatible 
and torsion-free).
Within specific contexts in general relativity, there are often physical 
reasons to transport specific vectors by specialized transport equations.
For example, along an accelerating worldline, it is often useful to carry
vectors using Fermi-Walker transport (see, e.g., \cite{MTW}).
Similarly, for a spinning point particle, its 4-momentum and angular-momentum 
tensor are jointly transported through the coupled Mathisson-Papapetrou 
equations \cite{Mathisson1937,Papapetrou1951} (the dual of these equations,
the Killing transport equations, also transport a vector and antisymmetric
tensor in a related way---see, e.g., \cite{Harte2008}).
In \cite{FlanaganNichols2014}, one of the authors and a collaborator 
introduced an affine transport equation for vectors that proved useful for 
measuring physical effects related to the gravitational-wave memory and for 
transporting a type of special-relativistic linear and angular momentum in 
general relativity.
We review the definition of this transport equation and describe some of its
properties in the next subsection.

\subsection{Affine transport equations of \cite{FlanaganNichols2014}}

The aim of \cite{FlanaganNichols2014} was to define an operational method
by which observers in asymptotically flat spacetimes could measure the
linear and angular momentum of the spacetime geometry from the spacetime
curvature and its derivatives in the viscinity of an observer's timelike
worldline.
The observers could then compare their measured values of linear and angular 
momentum by using a specific transport equation (which has the form of 
an affine map between the tangent spaces along a curve that connects two 
points along the two different worldlines).
In the case of spacetimes that are stationary, followed by a burst of 
gravitational waves with memory, and then stationary again, there was
observer dependence in the measured angular momentum that was a consequence
of the gravitational waves' memory.
Since the memory is related to the supertranslation degree of freedom in
the Bondi-Metzner-Sachs (BMS) group \cite{Bondi1962,Sachs1962}, the 
measurement and transport procedure was able to probe aspects of the 
angular-momentum ambiguity in general relativity.

The covariant method for transporting angular momentum and measuring 
gravitational-wave memory effects was based on a system of differential 
equations called ``affine transport'' in 
\cite{FlanaganNichols2014}.
Given a vector $\xi^a$ along a curve $x(\lambda)$, the affine transport of
$\xi^a$ was defined by 
\begin{equation}
\label{eq:AffineTrans}
\dot x^b\nabla_b\xi^a=\alpha\dot x^a\,,
\end{equation}
where $\dot x^a=dx^a/d\lambda$ is the tangent to the curve.
For simplicity, we will assume $\alpha=1$ throughout the remainder of this 
paper.
The solution $\xi^a$ at $x=x(\lambda)$, given the vector $\xi^{a'}$ at an 
initial point $x'=x(0)$, can be written as a sum of homogeneous and 
inhomogeneous parts,
\begin{equation}
\xi^a=\Lambda^a{}_{a'}\xi^{a'}+\Delta\xi^a\,.
\label{eq:AffineSol}
\end{equation}
Here $\Lambda^a{}_{a'}(\lambda)$ is the parallel transport map along the curve 
from $x'$ to $x$, which satisfies
\be
\dot x^b\nabla_b {\Lambda^a}_{a'}=0 \, ,\qqq 
{\Lambda^a}_{a'}(0)={\delta^a}_{a'}\,,
\ee
and $\Delta\xi^a(\lambda)$ is the inhomogeneous part of the solution, satisfying
\be
\dot x^b\nabla_b\Delta\xi^a=\dot x^a \,,\qqq \Delta\xi^a(0)=0\,.
\ee
The solution takes the form of an affine map between the two tangent spaces 
at $x$ and $x'$, which was the motivation for the name affine transport.

Only after \cite{FlanaganNichols2014} was written did it come to the attention
of the authors of \cite{FlanaganNichols2014} that the transport equations 
(\ref{eq:AffineTrans}) and their solution (\ref{eq:AffineSol}) are related
to other aspects of general relativity and differential geometry.
The vector $\Delta\xi^a$ also appears as the development of a curve on a 
manifold into the affine tangent space at the curve's starting point.
This is sometimes equivalently described as rolling the manifold along the 
initial tangent space without slipping or twisting 
\cite{KobayashiNomizu1963,Sharpe1997,ChitourKokkonen2012}.
More specifically, a vector $\Delta\xi^{a'}$ at $x'$ is equivalent to the
displacement vector in the initial affine tangent space that points between 
the initial and final values of the rolling (or developing) curve, and 
$\Delta\xi^a=\Lambda^a{}_{a'}\Delta\xi^{a'}$ is its parallel transport along 
the curve in the manifold from $x'$ to $x$.
In flat spacetime, $\Delta\xi^a$ is the net displacement vector from $x'$ to 
$x$, and in curved spacetime, $\Delta\xi^a$ provides a curve-dependent notion
of a displacement vector between the two points.

In addition, there is another construction in which the transport equation 
(\ref{eq:AffineTrans}) appears, as we now describe.
As a slight generalization of the linear bundle of orthonormal frames, one
can consider the affine frame bundle, in which the frame field is defined in
an affine space and consequently an additional vector defining the origin of 
this affine tangent space is also required.
A connection on this affine frame bundle prescribes that the vector is 
transported via (\ref{eq:AffineTrans}) (see, e.g., \cite{Tod1994} for a 
discussion of this in the relativity literature and for its use in 
understanding spacetimes with conical deficits).
It has long been known \cite{KobayashiNomizu1963} that there is a one-to-one
mapping between connections on the affine frame bundle and connections with 
torsion on the linear frame bundle.
In fact, for a connection on the affine frame bundle, the curvature of this
connection contains both the usual Riemann curvature of the linear frame 
bundle and the torsion as the curvature associated with the part of the 
connection that determines the transport of the additional vector 
(see, e.g., \cite{Tod1994}).
Thus, the holonomy for the affine frame bundle has a ``translational'' part
related to the torsion and a ``rotational'' part associated with the Riemann
curvature.

We will work exclusively with the metric-compatible, torsion-free derivative
$\nabla_a$ in this paper, however.
Within this context, it is not immediately clear how the holonomy found from
solving the affine transport equation around a closed curve (which was 
called a ``generalized holonomy'' in \cite{FlanaganNichols2014}) will behave
in the limit of an infinitesimal loop.
For large loops the inhomogeneous part of the solution is known to be 
nonvanishing because of nonlocal effects of spacetime curvature 
\cite{Petti1986,Tod1994,FlanaganNichols2014}.
In these three references, the inhomogeneous part of the solution has direct
physical relevance, because it can be used to find spacetimes that contain 
``torsion without torsion,'' understand certain spinning cosmic-string 
spacetimes, and measure gravitational-wave memory effects and observer 
dependence in angular momentum, respectively.

Our focus in this paper, therefore, will be to investigate the local 
geometrical properties of the affine transport equation (\ref{eq:AffineTrans}) 
and its holonomy around infinitesimal loops for torsion-free connections.
Our aim will be to find the relevant physical information that can be 
extracted from these local holonomies.
When we compute this generalized holonomy around small (contractible) loops in 
a generic (smooth) pseudo-Riemannian manifold, we find that the inhomogeneous
solution is a higher-order effect in the size of the loop for a torsion-free
connection.
In addition, the inhomogeneous part depends on just the Riemann tensor so 
that it contains the same physical data as the linear holonomy associated
with parallel transport.
Furthermore, by carefully defining the loop and using the methods of covariant 
bitensor calculus (see, e.g., \cite{Synge,DeWittBrehme,PoissonReview}), we 
can compute higher-order corrections to the generalized holonomy in the size
of the loop.
It remains the case that the homogeneous and inhomogeneous solutions contain
similar information about the gradient of the Riemann tensor, but the 
inhomogenous solution scales more rapidly with the loop's size.
We next provide an overview of the methods, results, and organization of this
paper in the next subsection.

\subsection{Summary of the results of this paper}
\label{sec:Results}

\begin{figure}[htb]
\begin{center}
\includegraphics[scale=.6]{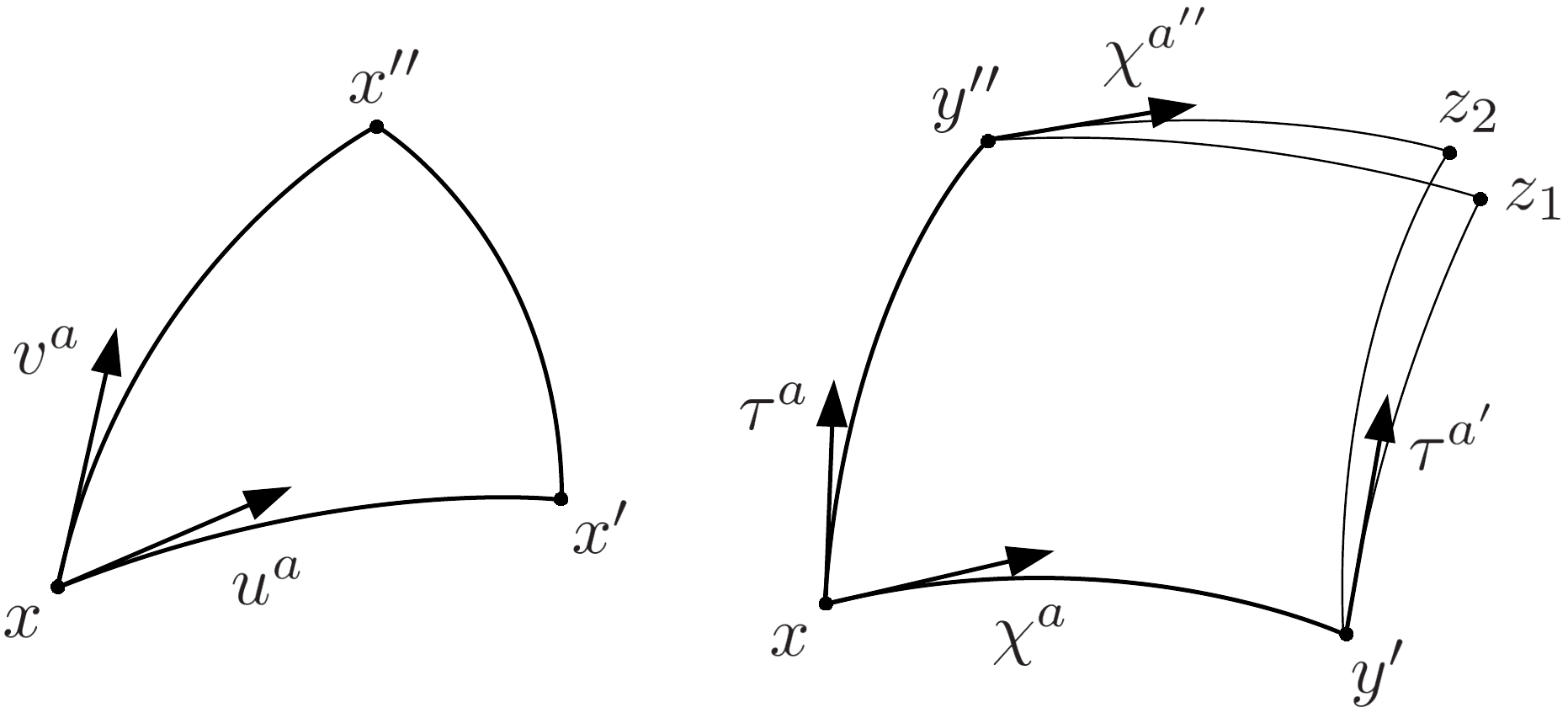}
\end{center}
\caption{\label{fig:intro}
\emph{Left}:  What we call the ``$u,v$ triangle'', denoted by 
$\triangle_{u,v}$, which is traversed counterclockwise (the $x $$\to$$ x'$$\to$$ x''$$\to$$ x$ direction).
The generalized holonomy of this loop is given by (\ref{TriLambda}) and 
(\ref{TriDeltaxi}).
The points $x'$ and $x''$ are the images of the exponential maps of $u^a$ 
and $v^a$ at $x$.
\emph{Right}: There are two possible parallelogramoid loops that can be
defined by a pair of vectors $\chi^a$ and $\tau^a$ at $x$, which we label by
\mbox{$x$$\to$$y'$$\to$$z_1$$\to$$y''$$\to$$x$} and $x$$\to$$ y'$$\to$$z_2$$\to$$y''$$\to$$ x$.
The points $y'$ and $y''$ are obtained through the exponential maps of 
$\chi^a$ and $\tau^a$ at $x$, respectively.  
The vector $\tau^{a'}$ at $y'$ is the parallel transport (along the $x\to y'$
geodesic) of $\tau^a$ at $x$, and the point $z_1$ comes from the
exponential map of $\tau^{a'}$ at $y'$; there is then a unique geodesic
linking $z_1$ to $y''$ and completing the first parallelogramoid.  
The vector $\chi^{a''}$ at $y''$ is the parallel transport (along the $x\to y''$ geodesic) of $\chi^a$ at $x$, and the point $z_2$ is the result of the 
exponential map of $\chi^{a''}$ at $y''$; there is then a unique geodesic that links $y'$ to $z_2$, thereby closing the second parallelogramoid. 
Through third order in distance, the generalized holonomy is the same around
either loop, and (\ref{LCpLambda}) and (\ref{LCpDeltaxi}) give the holonomy of
what we call the $\chi,\tau$ parallelogramoid loop,
$\diamondsuit_{\chi,\tau}$.
}
\end{figure}

By specializing the solution in (\ref{eq:AffineSol}) to a closed curve 
beginning and ending at a point $x$, we note that the solution to the affine 
transport equation (\ref{eq:AffineTrans}) around the loop defines an affine 
map on the tangent space at $x$.
It takes an initial vector $\xi_0^a$ at $x$ and returns a final vector 
$\xi^a$ at $x$:
\be\label{genhol}
\dot x^b\nabla_b\xi^a=\dot x^a
\qq\Rightarrow\qq
\xi^a=\Lambda^a{}_b\xi_0^b+\Delta\xi^a\, .
\ee
The linear map $\Lambda^a{}_b$ is the linear holonomy associated with the
metric-compatible, torsion-free derivative operator $\nabla_a$, and the 
vector $\Delta\xi^a$ is the inhomogeneous contribution to 
generalized holonomy, $(\Lambda^a{}_b,\Delta\xi^a)$.
We compute the generalized holonomy along the curves in Fig.~\ref{fig:intro},
and we list the results of our calculation in equations
(\ref{TriLambda})--(\ref{LCpDeltaxi}).

We first compute the generalized holonomy around a small geodesic triangle 
(on the left in Fig.~\ref{fig:intro}) defined by three points $x$, $x'$ and $x''$.
We assume that these points are within a convex normal neighborhood of one
another so that there exist unique geodesic segments connecting them.  
The solution can be expressed in terms of the two vectors $u^a$ and $v^a$ at 
$x$ that yield the points $x'$ and $x''$ under the exponential map.
For the loop followed counterclockwise ($x\to x'\to x''\to x$)---the 
``$u,v$ triangle'' or $\triangle_{u,v}$, for short---we show that the 
linear holonomy associated with parallel transport is
\be\label{TriLambda}
\Lambda^a{}_b(\triangle_{u,v})=\delta^a{}_b+\frac{1}{2}R^a{}_{bcd}v^c u^d+\frac{1}{6}R^a{}_{bcd;e}v^cu^d(v^e+u^e)+O(4)\,,
\ee
and the inhomogeneous solution is
\be\label{TriDeltaxi}
\Delta\xi^a(\triangle_{u,v})=\frac{1}{6}R^a{}_{bcd}(v^b+u^b)v^cu^d+O(4)\,.
\ee
Here $O(n)$ stands for terms with $n$ or more factors of the vectors 
$u^a$ and $v^a$ (i.e., $n$ powers of distance).
The result (\ref{TriDeltaxi}) shows that the leading-order inhomogeneous part 
of the generalized holonomy scales as the area of the triangle to the 
three-halves power (three powers of distance) times the Riemann tensor.
Because we computed (\ref{TriDeltaxi}) using the torsion-free, metric-compatible
derivative $\nabla_a$, the solution depends just upon the Riemann tensor and
is a higher-order effect.
We also give an exact series solution to all orders in distance (written in
terms of usual two-point coincidence limits of the parallel propagator), and
we present explicit results through fourth-order in distance in 
Appendix~\ref{app:fourthorder}.

We next consider the holonomies around small ``Levi-Civita parallelogramoids'' 
\cite{Cartan2001}, the quadrilaterals formed from geodesic segments that are 
the closest approximation in curved space to a flat-space parallelogram.  
As described in Fig.~\ref{fig:intro} and Sec.~\ref{sec:parallelogramoids}, we can use 
a pair of vectors $\chi^a$ and $\tau^a$ at a point $x$ to define two distinct 
parallelogramoid loops starting and ending at $x$.  
However, both loops in Fig.~\ref{fig:intro}, when traversed in the counterclockwise
direction from $x$, have the same generalized holonomy through third order in 
distance.
It is given by
\be\label{LCpLambda}
\Lambda^a{}_b(\diamondsuit_{\chi,\tau})=\delta^a{}_b+R^a{}_{bcd}\tau^c \chi^d+\frac{1}{2}R^a{}_{bcd;e}\tau^c\chi^d(\tau^e+\chi^e)+O(4)\,,
\ee
and
\be\label{LCpDeltaxi}
\Delta\xi^a(\diamondsuit_{\chi,\tau})=\frac{1}{2}R^a{}_{bcd}(\tau^b+\chi^b)\tau^c\chi^d+O(4)\,.
\ee
We show how the parallelogramoid solution can be obtained from a composition 
of the solutions for two triangles (which, through this order, is additive).

We now outline how we arrive at these results.
In Sec.~\ref{sec:MathPrelim}, we describe some basic mathematical results
that will be needed to derive the generalized holonomy around a triangle
and parallelogramoid.
Specifically, in Sec.~\ref{sec:AffineGeo}, we give the solution to the affine 
transport equation along a geodesic segment in terms of fundamental bitensors
(the parallel propagator and derivatives of Sygne's world function).
Section~\ref{sec:GeoTriangle} covers the mathematical framework for defining the
geodesic triangles.
The framework is similar to that used in \cite{Synge1931,Synge} to derive the 
curvature corrections to the law of cosines (a result which we reproduce 
below).
Section~\ref{sec:GenHolHom} contains a derivation of the linear holonomy around the 
triangle and discussion about how this method relates to other procedures
for computing the holonomy (e.g., using a path-ordered integral).
Section~\ref{sec:GenHolInhom} gives the inhomogeneous contribution to the 
generalized holonomy.
We treat the generalized holonomy of the parallelogramoid in 
Sec.~\ref{sec:parallelogramoids}, and in Sec.~\ref{sec:Implications} we 
discuss the implications of these calculations for the program described 
in~\cite{FlanaganNichols2014}.
We conclude in Sec.~\ref{sec:Conclusion}.
Appendix~\ref{app:fourthorder} contains fourth-order terms for the generalized 
holonomy of the geodesic triangle.

\section{\label{sec:MathPrelim} Mathematical preliminaries}

Because the formalism of covariant bitensors is carefully explained in the 
review paper \cite{PoissonReview}, we will refer the reader to that resource
for more background and detail on bitensor calculus.
We generally adopt the notation of \cite{PoissonReview}, except that we use 
Latin rather than Greek tensor indices, and we often interchange the role of
the primed and unprimed indices (corresponding to tensor indices in the 
tangent spaces of different spacetime points) relative to \cite{PoissonReview}.
Thus, for two spacetime points $x$ and $x'$ connected by a geodesic, we will
use $\sigma(x,x')$ to denote Synge's world function (half the squared proper
distance along the geodesic between the points) and ${g^{a'}}_a$ to denote 
the parallel propagator (which was written as ${\Lambda^{a'}}_a$ in the 
introduction).
We will use semicolons preceding indices to denote the covariant derivative
operator (e.g., $\sigma_{;a}$ or $\sigma_{;a'}$) and square brackets around
quantities to denote coincidence limits (e.g., $[{g^{a'}}_b] = {\delta^a}_b$).

\subsection{Affine transport along a geodesic segment}
\label{sec:AffineGeo}

Consider an affinely parametrized geodesic $x'(\lambda)$ with tangent 
$u^{a'}(\lambda)$,
\be
u^{a'}=\frac{dx^{a'}}{d\lambda}\, ,\qqq
\frac{Du^{a'}}{D\lambda}=u^{b'}\nabla_{b'}u^{a'}=0 \, ,
\ee
and let $x=x'(0)$ be a fixed initial point on the geodesic, where the tangent 
is $u^a$.  
The affine transport equation,
\begin{equation}\label{geoaff}
u^{b'}\nabla_{b'}\xi^{a'}=u^{a'}\,,
\end{equation}
has a formal solution along the (assumed unique) geodesic connecting $x$ to 
$x'(\lambda)$, which in the language of bitensor calculus 
\cite{Synge,DeWittBrehme,PoissonReview} is given by
\begin{equation}\label{geoaffsol}
\xi^{a'}={g^{a'}}_{a}(x,x')\,\xi^{a}+\sigma^{;a'}(x,x')\,.
\end{equation}
Here $\xi^{a'}$ is the solution at $x'$, $\xi^a$ is the initial value at $x$, 
${g^{a'}}_{a}(x,x')$ is the parallel propagator, and $\sigma^{;a'}(x,x')$ is 
the covariant derivative at $x'$ of Synge's world function $\sigma(x,x')$.

\begin{figure}[h]
\begin{center}
\includegraphics[scale=.6]{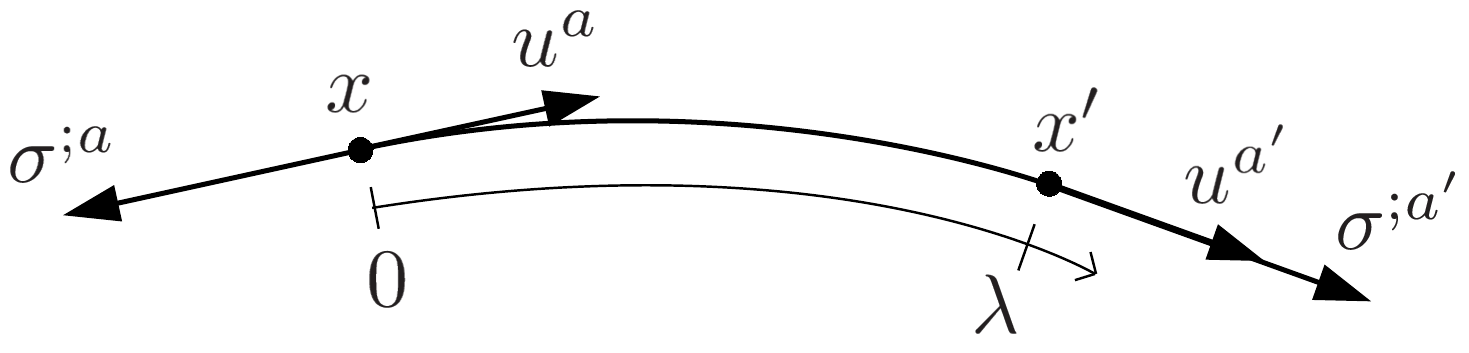}
\end{center}
\caption{\label{fig:Geodesic}
The affinely parametrized geodesic $x'(\lambda)$ with initial point 
$x=x'(0)$.  
The tangent is parallel transported from $u^a$ at $x$ to $u^{a'}$at $x'$.
The derivatives of the world function both point outward from the geodesic 
segment: $\sigma^{;a}=-\lambda u^a$ and $\sigma^{;a'}=\lambda u^{a'}$.}
\end{figure}

That (\ref{geoaffsol}) satisfies (\ref{geoaff}) follows from the following 
properties:
The world function is related to the tangents and the affine parameter 
interval by
\be\label{las}
\sigma(x,x')=\frac{1}{2}\lambda^2u^2\,,\qqq 
\sigma^{;a}=-\lambda u^a\,,\qqq \sigma^{;a'}=\lambda u^{a'}\,,
\ee
and it satisfies
\be\label{seco}
\sigma^{;b}{\sigma^{;a}}_b=\sigma^{;a}\,,\qqq 
\sigma^{;b'}{\sigma^{;a'}}_{b'}=\sigma^{;a'} \, .
\ee
Dividing the second equation of (\ref{seco}) by $\lambda$ and using the last
equation of (\ref{las}) shows that the second term in (\ref{geoaffsol}) is 
the inhomogeneous (particular) solution to (\ref{geoaff}).  
That the first term of (\ref{geoaffsol}) is the homogeneous solution follows 
from the second of the identities
\be\label{defpp}
\sigma^{;b} {g^{a'}}_{a;b}=0\,,\qqq \sigma^{;b'} {g^{a'}}_{a;b'}=0\,,
\ee
and the condition ${g^{a'}}_a\to{\delta^{a'}}_a$ as $x'\to x$ that defines 
the parallel propagator.
Also note that while the tangent is parallel transported, the world function
derivatives are minus the parallel transports of each other:
\be\label{tangsig}
u^{a'}={g^{a'}}_au^a\,,\qqq \sigma^{;a'}=-{g^{a'}}_a\sigma^{;a}\,.
\ee
The above properties will be used often throughout the remainder
of the paper.

\subsection{Geodesic triangles}
\label{sec:GeoTriangle}

\begin{figure}[htb]
\begin{center}
\includegraphics[scale=.6]{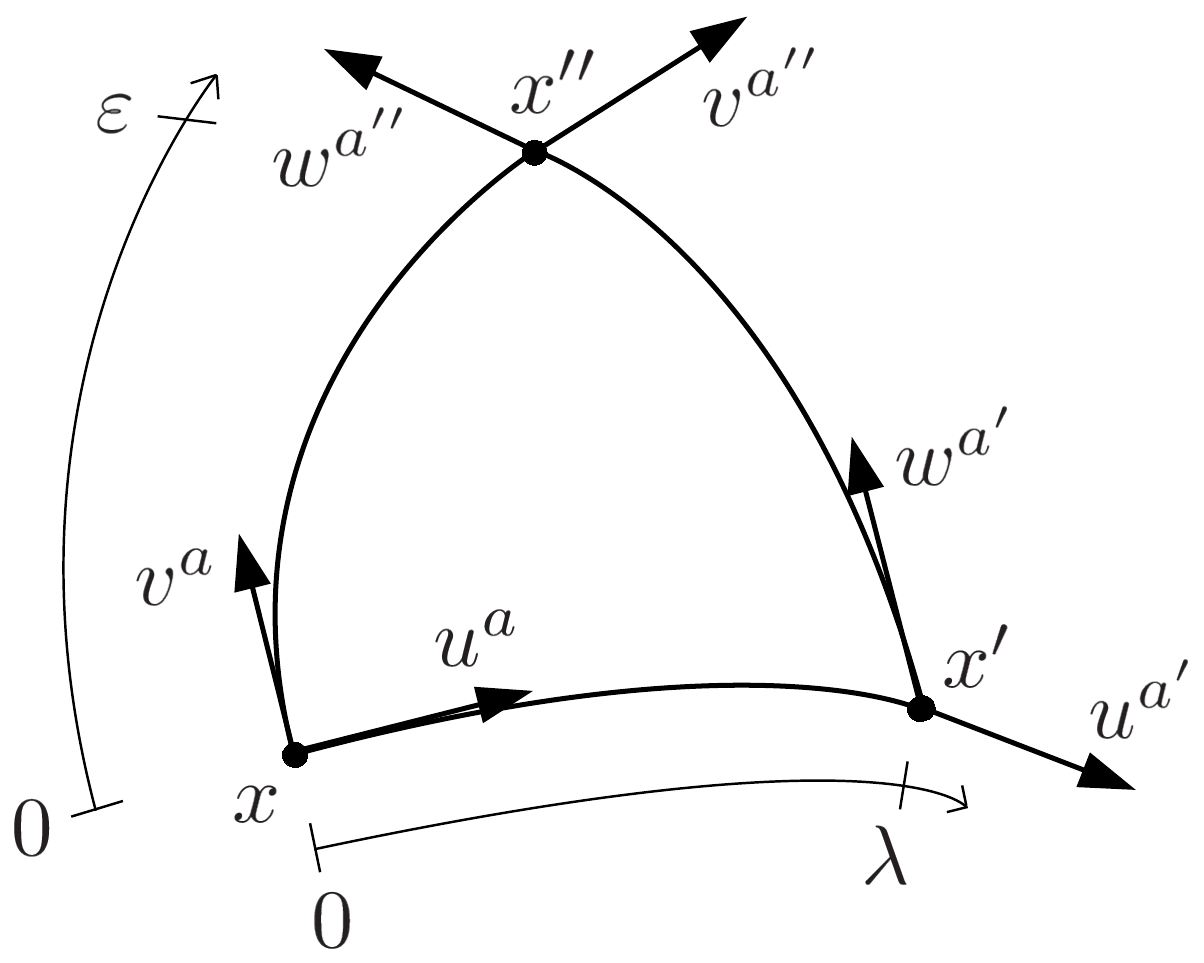}
\end{center}
\caption{\label{fig:triangle}
The geodesic triangle associated with the three points $x$, $x'$, and $x''$.
We find it convenient to treat the triangle as a 
function of a fixed point $x$ and fixed initial tangents $u^a$ and $v^a$ at 
$x$ with two affine parameters $\lambda$ and $\ve$ that parameterize two
of the geodesic legs.  The tangents are parallel transported along the legs: 
$u^a$ at $x$ to $u^{a'}$ at $x'$, $v^a$ at $x$ to $v^{a''}$ at $x''$, and 
$w^{a'}$ at $x'$ to $w^{a''}$ at $x''$.}
\end{figure}

We now describe our framework for defining geodesic triangles 
(see Fig.~\ref{fig:triangle}).
The somewhat lengthy and detailed definition of the triangle is necessary to 
compute $\Delta \xi^a$ without ambiguity (and also to obtain any higher-order 
corrections to both ${\Lambda^a}_b$ and $\Delta \xi^a$).

We start at a fixed base point $x$ with two vectors $u^a$ and $v^a$,
and we then follow the geodesics with initial tangents $u^a$ and $v^a$ 
for affine parameter intervals $\lambda$ and $\ve$ to reach the points $x'$ 
and $x''$, respectively.
As in (\ref{las}), the tangents are related to the world-function derivatives 
by
\be\label{uvle}
\lambda u^a=-\sigma^{;a}(x,x'),\qqq \ve v^a=-\sigma^{;a}(x,x'')\, .
\ee
This defines affinely parametrized geodesics $x'(\lambda)$ and $x''(\ve)$ 
emanating from $x$.  
The tangents to these geodesics at $x'$ and $x''$ are given by
\be\label{uap}
\lambda u^{a'}=\sigma^{;a'}(x,x'),\qqq \ve v^{a''}=\sigma^{;a''}(x,x'')\,,
\ee
which are parallel transports of (\ref{uvle}) [cf.~(\ref{tangsig})].
We assume there is then a unique geodesic segment connecting $x'$ to $x''$.
Its tangents are denoted by $w^{a'}$ at $x'$ and $w^{a''}$ at $x''$, and are 
assumed to be normalized so that the affine parameter interval from $x'$ 
to $x''$ is 1.
In terms of derivatives of the world function, they are given by
\be
w^{a'}=-\sigma^{;a'}(x',x''),\qqq w^{a''}=\sigma^{;a''}(x',x'')\,,
\ee
and they are related to each other by parallel transport along the 
geodesic connecting $x'$ and $x''$.

In our calculations below, we will fix the base point $x$ and the vectors 
$u^a$ and $v^a$ at $x$, and we will vary the affine parameters $\lambda$ and
$\ve$.
From this perspective, the points $x'$ and $x''$ vary along the fixed
geodesics determined by $u^a$ and $v^a$ at $x$.
Quantities expressible as functions of $x'$ and $x''$, such as 
$w^{a'}=-\sigma^{;a'}(x',x'')$ or ${g^{a''}}_{a'}(x',x'')$, can then be 
expressed as functions of $\lambda$ and $\ve$, and can be differentiated 
according to
\be\label{ddldde}
\frac{D}{D\lambda}=u^{a'}\nabla_{a'}\,,\qqq 
\frac{D}{D\ve}=v^{a''}\nabla_{a''}\,.
\ee
Note that these two derivatives commute, because $\nabla_{a'}$ and 
$\nabla_{a''}$ commute, and because $u^{a'}$ is independent of $\ve$ and 
$v^{a''}$ is independent of $\lambda$.  
Also note that the quantities $u^{a'}$ and $g^{a'}{}_a(x,x')$ depend only on
$\lambda$; $v^{a''}$ and $g^a{}_{a''}(x,x'')$ depend only on $\ve$; and  
$w^{a'}$, $w^{a''}$, and $g^{a''}{}_{a'}(x',x'')$ depend on both 
$\lambda$ and $\ve$.
These are all of the quantities necessary to define the generalized holonomy 
around the triangle.

\section{Generalized holonomy for geodesic triangles}

\subsection{Linear part of the holonomy}
\label{sec:GenHolHom}

We now turn to the calculation of the holonomy of parallel transport around
the geodesic triangle loop of the previous section.
When following the loop counterclockwise ($x$$\to$$ x'$$\to$$ x''$$\to$$ x$), 
the holonomy of parallel transport is given by
\be\label{triho}
{\Lambda^a}_b(\triangle_{u,v})={g^a}_{a''} \,{g^{a''}}_{b'} \,{g^{b'}}_{b} =
\sum_{m,n=0}^\infty \frac{\lambda^m\ve^n}{m!n!}{\Lambda^a}_{b(m,n)}\,.
\ee
We leave out the arguments of the parallel propagators, because they can
be understood from their indices.
Assuming $x$, $u^a$, and $v^a$ fixed, while $\lambda$ and $\ve$ vary to 
change the locations of $x'$ and $x''$, we write the holonomy tensor as a 
covariant Taylor series in $\lambda$ and $\ve$ (the second equality).
The coefficients ${\Lambda^a}_{b(m,n)}$ are constant tensors at $x$ that will 
depend on $u^a$, $v^a$, and the local geometry at $x$.

We briefly digress to discuss the series solution for the holonomy in 
(\ref{triho}).
There is also a different formal solution for the holonomy in terms of 
path-ordered integrals of the frame components of the Riemann tensor along
a closed curve (see, e.g., \cite{Vickers1987}).
Unlike the solution in (\ref{triho}), the path-ordered-integral solution is 
typically given for an arbitrary curve (not necessarily a triangle formed 
from geodesics) and it is expressed in terms of integrals over all the points
along the curve (not necessarily just the initial point $x$).
If the path-ordered-integral solution were specialized to a curve that traces
out a small geodesic triangle, and the components of the Riemann tensor along 
the curve were expanded in terms of covariant bitensors around the point $x$, 
then the path-ordered-integral solution should reduce to (\ref{triho}).
Because the series solution in (\ref{triho}) generalizes quite straightforwardly
to computing the inhomogeneous solution, we choose to use this method rather
than the path-ordered integral hereafter.

We can calculate the coefficients ${\Lambda^a}_{b(m,n)}$ by repeatedly 
differentiating (\ref{triho}), using the operators 
$\frac{D}{D\lambda}=u^{a'}\nabla_{a'}$ and 
$\frac{D}{D\ve}=v^{a''}\nabla_{a''}$ of (\ref{ddldde}).
When doing so, we will frequently use the identities
\be\label{thanx}
u^{a'}{}_{;b'}u^{b'}=0=v^{a''}{}_{;b''}v^{b''}\, ,\qqq 
{g^{a'}}_{b;c'}u^{c'}=0={g^a}_{a'';c''}v^{c''} \,,
\ee
which is a restatement of the geodesic equations for $x'(\lambda)$ and 
$x''(\ve)$ and additionally, one of the defining properties of the parallel 
propagators [cf. (\ref{defpp}) and (\ref{uap})]. 
The $(0,0)$ coefficient is given by the limit of (\ref{triho}) as 
$\lambda\to0$ and $\ve\to0$ and is the identity map:
\be
\Lambda^a{}_{b(0,0)}=\delta^a{}_b\,.
\ee
Acting on (\ref{triho}) with $m$ $\lambda$-derivatives and $n$ 
$\ve$-derivatives and using (\ref{thanx}), we find
\bea\label{derivs}
\left(\frac{D}{D\lambda}\right)^m\left(\frac{D}{D\ve}\right)^n\Lambda^a{}_b&=&g^a{}_{a''}g^{a''}{}_{b';c'_1\ldots c'_md''_1\ldots d''_n} u^{c'_1}\ldots u^{c'_m}v^{d''_1}\ldots v^{d''_n}g^{b'}{}_b
\nnm\\
&=&\Lambda^a{}_{b(m,n)}+O(\lambda)+O(\epsilon)\,.
\eea
Taking the $\ve\to 0$ ($x''\to x$) limit of this equation yields
\be\label{firstlimit}
\Lambda^a{}_{b(m,n)}+O(\lambda)=g^{a}{}_{b';c'_1\ldots c'_md_1\ldots d_n} u^{c'_1}\ldots u^{c'_m}v^{d_1}\ldots v^{d_n}g^{b'}{}_b\,,
\ee
and then taking the $\lambda\to 0$ ($x'\to x$) limit yields the expansion 
coefficients in terms of usual two-point coincidence limits 
\cite{PoissonReview},
\be\label{Lambdacoeff}
\Lambda^a{}_{b(m,n)}=\Big[g^{a}{}_{b';c'_1\ldots c'_md_1\ldots d_n}\Big]_{x'\to x} u^{c_1}\ldots u^{c_m}v^{d_1}\ldots v^{d_n}\,.
\ee
Recall that in this notation, the coincidence limit turns primed indices
associated with the tangent space at $x'$ to unprimed indices associated 
with those at $x$.
Note that, had we taken the limits in the opposite order, we would have obtained
\bea
\Lambda^a{}_{b(m,n)}&=&\Big[g^{a''}{}_{b;c_1\ldots c_md''_1\ldots d''_n}\Big]_{x''\to x} u^{c_1}\ldots u^{c_m}v^{d_1}\ldots v^{d_n}
\\\nnm
&=&\Big[g^{a'}{}_{b;c_1\ldots c_md'_1\ldots d'_n}\Big]_{x'\to x} u^{c_1}\ldots u^{c_m}v^{d_1}\ldots v^{d_n}\,,
\eea
where the second line has inconsequentially renamed $x''$ to $x'$.  
We see that consistency requires the identity
\be\label{theidentity}
\Big[g^{a}{}_{b';(c'_1\ldots c'_m)(d_1\ldots d_n)}\Big]_{x'\to x}=\Big[g^{a'}{}_{b;(c_1\ldots c_m)(d'_1\ldots d'_n)}\Big]_{x'\to x}\,.
\ee
This identity actually holds without the symmetrizations and is a special 
case of the more general identity
\be\label{Jordan}
\Big[T_{a'_1\ldots a'_mb_1\ldots b_n}(x,x')\Big]_{x'\to x}=\Big[T_{a_1\ldots a_mb'_1\ldots b'_n}(x',x)\Big]_{x'\to x}\,,
\ee
which holds for any bitensor $T$ with a well-defined coincidence limit and 
which is a consequence of the independence of the path of approach to 
coincidence.\footnote{We thank Jordan Moxon for clarifying this
point for us.}  
Thus, the two limits taken in (\ref{firstlimit}) and (\ref{Lambdacoeff}) commute
[like the two derivatives (\ref{derivs}) do].

The coincidence limits necessary to compute $\Lambda^a{}_b$ through third 
order, via the expression (\ref{Lambdacoeff}), are given by
$$
\Big[g^{a}{}_{b';c}\Big]=0=\Big[g^{a}{}_{b';c'}\Big]\,,
$$
$$
\Big[g^{a}{}_{b';cd}\Big]=-\frac{1}{2}R^a{}_{bcd}\,,\qq \Big[g^{a}{}_{b';cd'}\Big]=\frac{1}{2}R^a{}_{bcd}\,,
\qq \Big[g^{a}{}_{b';c'd'}\Big]=\frac{1}{2}R^a{}_{bcd}
$$
$$
\Big[g^{a}{}_{b';cde}\Big]=-\frac{2}{3}R^a{}_{bc(d;e)}\,,\qq \Big[g^{a}{}_{b';cde'}\Big]=-\frac{1}{3}R^a{}_{be(c;d)}\,,
$$
$$
\Big[g^{a}{}_{b';cd'e'}\Big]=\frac{1}{3}R^a{}_{bc(d;e)}\,,\qq \Big[g^{a}{}_{b';c'd'e'}\Big]=\frac{2}{3}R^a{}_{bc(d;e)}\,.
$$
The limits in the first two lines are well-known results 
\cite{Synge,PoissonReview}, and we discuss the computation of those in the 
last two lines in Appendix~\ref{app:fourthorder}.
Substituting these relations and (\ref{Lambdacoeff}) into (\ref{triho}) and 
setting $\lambda=\ve=1$ (or, equivalently, absorbing $\lambda$ into the 
definition of $u^a$ and $\ve$ into that of $v^a$), we obtain the holonomy 
of parallel transport, 
\be\label{trihores}
\Lambda^a{}_b(\triangle_{u,v})=\delta^a{}_b+\frac{1}{2}R^a{}_{bcd}v^cu^d+\frac{1}{6}R^a{}_{bcd;e}v^cu^d(v^e+u^e)+O(4)\,.
\ee
We list the fourth-order corrections to this result in 
Appendix~\ref{app:fourthorder}.

The term in (\ref{trihores}) equal to $\frac{1}{2}R^a{}_{bcd}v^cu^d$ is half
the value that appears in many textbook derivations of the holonomy around
an ``infinitesimal parallelogram'' spanned by two vectors $v^a$ and $u^a$ 
(see, e.g., \cite{Nakahara2003}).
This is not surprising because the geodesic triangle has half its area.
We were unable to find an equivalent calculation in the literature against 
which to check the third-order term in this series, 
$\frac{1}{6}R^a{}_{bcd;e}v^cu^d(v^e+u^e)$.
It is possible that the result is not new, however.
This third-order term provides a quantitative estimate of the error in the
leading-order expression for ${\Lambda^a}_b$ for finite $\ve$ and $\lambda$.
It also enters at the same order in this expansion as the inhomogeneous part 
of the generalized holonomy, as we show next.

\subsection{Inhomogeneous part of the holonomy}
\label{sec:GenHolInhom}

Next, consider the inhomogeneous part $\Delta\xi^a$ of the generalized holonomy around the triangle.  
Under the same assumptions as in the calculation of the linear holonomy, we 
can write the exact solution by composing the solution (\ref{geoaffsol}) three
times along each leg.
Specifically, starting with $\xi^a=0$ at $x$, we obtain $\sigma^{a'}(x,x')$ 
at $x'$; then we parallel transport that to $x''$ and add 
$\sigma^{a''}(x'',x')$; finally, we parallel transport that to $x$ and add 
$\sigma^a(x,x'')$.
The net result is
\bea\label{triDx}
\Delta \xi^a&=&g^a{}_{a''}\left(g^{a''}{}_{a'}\sigma^{;a'}(x',x)+\sigma^{;a''}(x'',x')\right)+\sigma^{;a}(x,x'')
\nnm\\
&=&\Lambda^a{}_b\lambda u^b+g^a{}_{a''}w^{a''}-\ve v^a\,,
\eea
where the second line has used (\ref{triho}) and the definitions of 
Sec.~\ref{sec:GeoTriangle}.  
Having already computed $\Lambda^a{}_b$, we now need only to expand the 
quantity
\be\label{wtilde}
\tilde w^a\equiv g^a{}_{a''}w^{a''} = g^a{}_{a''}\sigma^{;a''}(x'',x') =
\sum_{m,n=0}^\infty \frac{\lambda^m\ve^n}{m!n!}\tilde w^a{}_{(m,n)}\,,
\ee
which is the tangent at $x''$ to the geodesic between $x'$ and $x''$ that has
been parallel transported back to $x$.  
Its coincidence limit is
\be
\tilde w^a{}_{(0,0)}=0\,.
\ee
The coefficients $\tilde w^a{}_{(m,n)}$ can be computed similarly to those
of the holonomy in (\ref{derivs}).
After using the relations (\ref{thanx}), derivatives of (\ref{wtilde}) have
the simple form
\bea
\left(\frac{D}{D\lambda}\right)^m\left(\frac{D}{D\ve}\right)^n\tilde w^a&=&g^a{}_{a''}\sigma^{;a''}{}_{c'_1\ldots c'_md''_1\ldots d''_n} u^{c'_1}\ldots u^{c'_m}v^{d''_1}\ldots v^{d''_n}
\nnm\\
&=&\tilde w^a{}_{(m,n)}+O(\lambda)+O(\epsilon)\,.
\eea
Taking the $\ve\to0$ ($x''\to x$) limit simplifies the expression to
\be
\tilde w^a{}_{(m,n)}+O(\lambda)=\sigma^{;a}{}_{c'_1\ldots c'_md_1\ldots d_n} u^{c'_1}\ldots u^{c'_m}v^{d_1}\ldots v^{d_n}\, ,
\ee
and taking the $\lambda\to0$ ($x'\to x$) limit leaves
\be\label{wcoeffs}
\tilde w^a{}_{(m,n)}= \Big[\sigma^{;a}{}_{c'_1\ldots c'_md_1\ldots d_n}\Big]_{x'\to x}u^{c_1}\ldots u^{c_m}v^{d_1}\ldots v^{d_n}\,.
\ee
As in the previous section, the two limits commute.  
To compute $\tilde w^a$ through fourth order, we first note that the
coincidence limits of all third derivatives of the world function vanish.
Then, the coincidence limits necessary to evaluate $\tilde w^a$ are given 
by \cite{Synge,PoissonReview}
$$
\Big[\sigma_{;ab}\Big]=g_{ab}\,,\qq\Big[\sigma_{;ab'}\Big]=-g_{ab}\,,\qq\Big[\sigma_{;a'b'}\Big]=g_{ab}\,,
$$
$$
\Big[\sigma_{;abcd}\Big]=S_{abcd}\,,\qq \Big[\sigma_{;abcd'}\Big]=-S_{abcd}\,,\qq \Big[\sigma_{;abc'd'}\Big]=S_{abcd}\,,
$$
$$
\Big[\sigma_{;ab'c'd'}\Big]=-S_{bcda}\,,\qq \Big[\sigma_{;a'b'c'd'}\Big]=S_{abcd}\,,
$$
where $S_{abcd}$ is the symmetrized Riemann tensor:
$$
S_{abcd}=S_{(ab)(cd)}=S_{(cd)(ab)}=-\frac{1}{3}\Big(R_{acbd}+R_{adbc}\Big)\,.
$$
Substituting these results into (\ref{wcoeffs}) and the series (\ref{wtilde}),
we obtain
\be\label{wres}
\tilde w^a= g^a{}_{a''}w^{a''} =v^a-u^a+\frac{1}{6}R^a{}_{bcd}\left(v^b-2u^b\right)v^cu^d+O(4)\, .
\ee
Putting this together with the results in (\ref{trihores}) and (\ref{triDx})
yields the inhomogeneous part of the generalized holonomy, 
\be\label{Dxres}
\Delta\xi^a(\triangle_{u,v})=\frac{1}{6}R^a{}_{bcd}(v^b+u^b)v^cu^d+O(4)\,.
\ee
Appendix~\ref{app:fourthorder} also contains the fourth-order corrections to 
this result, which give a quantitative error estimate on this leading result
for finite $\ve$ and $\lambda$.

Although we are unaware of another reference which has computed the 
inhomogeneous part of the generalized holonomy around a geodesic triangle
for a torsion-free derivative operator, we can compute a closely related 
quantity that appears in classical differential geometry as a check of our 
result.
We first note that an expansion similar to that above gives the tangent at 
$x'$ to the $x'$-$x''$ leg, parallel transported back to $x$, as
\be
g^a{}_{a'}w^{a'}=v^a-u^a+\frac{1}{6}R^a{}_{bcd}\left(u^b-2v^b\right)v^cu^d+O(4)\,,
\ee
which is simply (\ref{wres}) with $u^a\leftrightarrow v^a$ and an overall 
minus sign.
This result [or the result (\ref{wres}) for $g^a{}_{a''}w^{a''}$] 
contracted with itself provides the leading-order correction to the law of 
cosines in curved space \cite{Synge1931,Synge}.
Expressing the squared geodesic interval between $x'$ and $x''$ in terms of 
$u^a$ and $v^a$, we find this better-known result,
\be
w^2=(v-u)^2-\frac{1}{3}R_{abcd}v^au^bv^cu^d+O(5)\,.
\ee

With the inhomogeneous solution for the triangle, we can now compute the 
generalized holonomy around an infinitesimal parallelogramoid in the next
section.

\section{Generalized holonomy for parallelogramoids}
\label{sec:parallelogramoids}

There are two possible parallelogramoid loops which can be defined from two 
vectors $\chi^a$ and $\tau^a$ at a point $x$ (see the right half of 
Fig.~\ref{fig:intro} or Fig.~\ref{fig:chitaupsi} below).  
The points $y'$, $y''$, $z_1$, and $z_2$ are defined in terms of the vectors 
$\chi^a$ and $\tau^a$ at $x$ by
$$
\chi^a=-\sigma^{;a}(x,y')\,,\qq \tau^{a'}=g^{a'}{}_a(x,y')\,\tau^a=-\sigma^{;a'}(y',z_1)\,,
$$
$$
\tau^a=-\sigma^{;a}(x,y'')\,,\qq \chi^{a''}=g^{a''}{}_a(x,y'')\,\chi^a=-\sigma^{;a''}(y'',z_2)\,.
$$
We can define vectors $\psi^a_1$ and $\psi^a_2$ at $x$ to be the tangents to 
the ``diagonal'' geodesics connecting $x$ to $z_1$ and $z_2$, respectively, 
with unit affine parameter intervals along the segments.
They have the form
$$
\psi^a_1=-\sigma^{;a}(x,z_1)\,,\qqq \psi_2^a=-\sigma^{;a}(x,z_2)\,,
$$
and we can show that they are related to $\chi^a$ and $\tau^a$ by
\be\label{ok}
\psi^a_1+O(3)=\psi^a_2+O(3)=\chi^a+\tau^a\equiv \psi^a\,,
\ee
with the following argument:  
First, let us identify the $x$-$z_2$-$y''$ triangle with the $x$-$x'$-$x''$ 
triangle of Sec.~\ref{sec:GeoTriangle}.
The tangent vectors $\psi_2^a$, $\tau^a$, and $\chi^{a''}$ should then be 
identified with $u^a$, $v^a$, and $-w^{a''}$, respectively.
The result (\ref{wres}) then tells us that 
$-\chi^a=-g^a_{a''}\chi^{a''}=\tau^a-\psi_2^a+O(3)$.  
An analogous result holds for the $x$-$z_1$-$y''$ triangle, from which 
we obtain (\ref{ok}).
\begin{figure}[h]
\begin{center}
\includegraphics[scale=.6]{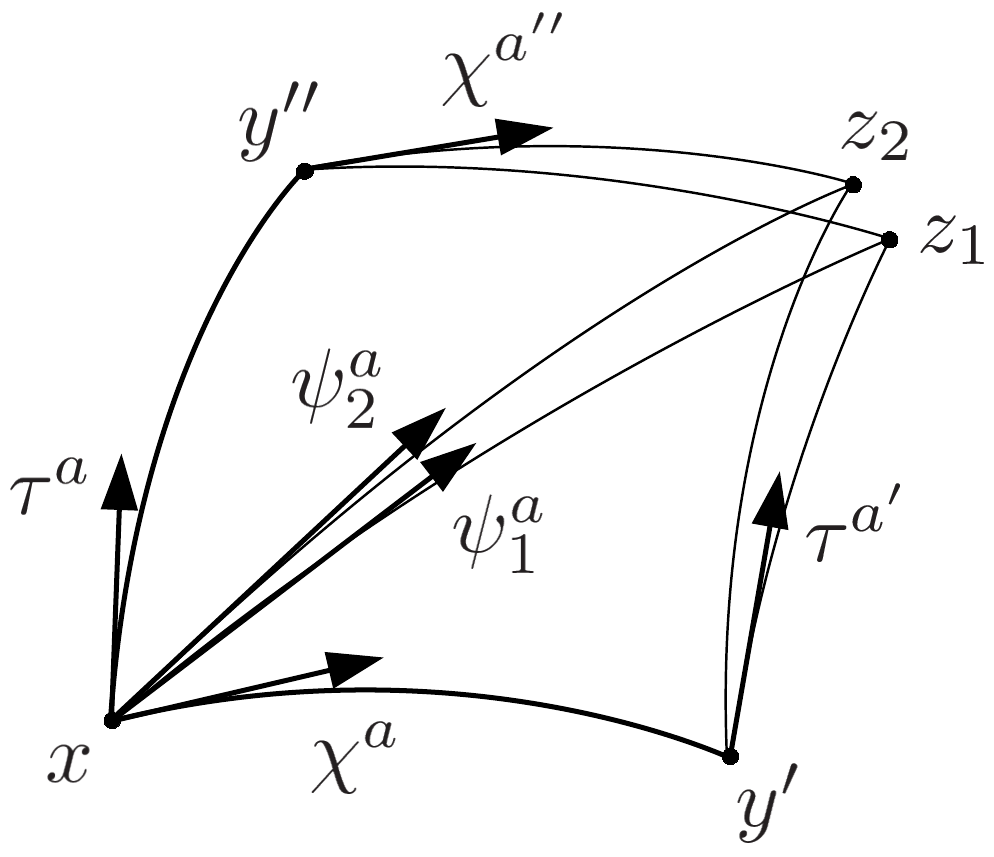}
\end{center}
\caption{\label{fig:chitaupsi}
The two parallelogramoid loops, 
\mbox{$x$$\to$$y'$$\to$$z_1$$\to$$y''$$\to$$x$} and $x$$\to$$ y'$$\to$$z_2$$\to$$y''$$\to$$ x$, as in Fig.~\ref{fig:intro}.  
Each parallelogramoid can be split into two triangles by defining a geodesic 
that runs along the diagonal from $x$ to $z_1$ or $z_2$.}
\end{figure}

The result of parallel or affine transport around the 
$x$$\to$$y'$$\to$$z_i$$\to$$y''$$\to$$x$ parallelogramoid loop (with 
$z_i=z_1$ or $z_2$) will be the same as the result of transport around the 
$x$$\to$$y'$$\to$$z_i$$\to$$x$ triangle loop followed by transport around the 
$x$$\to$$z_i$$\to$$y''$$\to$$x$ triangle loop, because transport along the 
last leg of the first triangle is the inverse of transport along the first 
leg of the second.
We find that, at this order, the same holonomies are obtained from either 
choice of the parallelogramoid loop.
Using the relations (\ref{trihores}) and (\ref{ok}), the linear holonomy 
$\Lambda^a{}_b$ is given by
\bea
\Lambda^a{}_b(\diamondsuit_{\chi,\tau})&=&\Lambda^a{}_c(\triangle_{\chi,\psi})\,\Lambda^c{}_b(\triangle_{\psi,\tau})+O(4)
\\\nnm
&=&\delta^a{}_b+R^a{}_{bcd}\tau^c \chi^d+\frac{1}{2}R^a{}_{bcd;e}\tau^c\chi^d(\tau^e+\chi^e)+O(4)\, .
\eea
Similarly, from (\ref{Dxres}) and (\ref{ok}), the inhomogeneous part of the 
generalized holonomy is given by
\bea
\Delta\xi^a(\diamondsuit_{\chi,\tau})&=&\Lambda^a{}_b(\triangle_{\psi,\tau})\,\Delta\xi^b(\triangle_{\chi,\psi})+\Delta\xi^a(\triangle_{\psi,\tau})+O(4)
\\\nnm
&=&\Delta\xi^a(\triangle_{\chi,\psi})+\Delta\xi^a(\triangle_{\psi,\tau})+O(4)\phantom{\Big|}
\\\nnm
&=&\frac{1}{2}R^a{}_{bcd}(\tau^b+\chi^b)\tau^c\chi^d+O(4)\,.
\eea
We see that, through this order, the generalized holonomy for the parallelogramoid(s)
can be found by simply adding the results ($\Lambda^a{}_b-\delta^a{}_b$ or $\Delta\xi^a$) from the generalized holonomies
of two appropriate triangles (which does not hold at higher orders).  We note that, through this order, the same result would also be found by traversing the $x$$\to$$y'$$\to$$z_1$$\to$$z_2$$\to$$y''$$\to$$x$ pentagon.

\section{Implications for Ref.~\cite{FlanaganNichols2014}}
\label{sec:Implications}

The results derived for the generalized holonomy around parallelogramoids 
above---and summarized in 
Eqs.~(\ref{LCpLambda})~and~(\ref{LCpDeltaxi})---may appear problematic for
the generalized holonomy of gravitational-wave spacetimes near future null 
infinity (the case considered in \cite{FlanaganNichols2014}).
Consider a linearized gravitational wave spacetime in transverse traceless
coordinates, for which the metric is given by
\begin{equation}
ds^2 = - dt^2 + (\delta_{ij} + r^{-1} h_{ij}^{\mathrm{TT}}) dx^i dx^j \, .
\end{equation}
The nonzero components of the Riemann tensor at leading order in $1/r$ 
are given by 
\begin{equation}
R_{titj} = \frac{\ddot h_{ij}^{\mathrm{TT}}}r + O(r^{-2}) \, ,
\end{equation}
where a dot was used to denote $\partial_t$.

Next, consider two observers at fixed large $r$ who are separated by 
$\delta x^i = r \delta \theta^i$, where 
$\sqrt{\delta \theta^i \delta \theta_i} = d\theta$ is small angle separating
the observers.
Impose that their 4-velocities are given by $\vec u = \vec \partial_t$, and 
allow an increment of time $\delta t$ to elapse along their worldlines.
The two worldlines and the separation $\delta x^i$ between the two observers
can be used to define a closed curve around which we can compute the 
generalized holonomy.
The results~(\ref{LCpLambda})~and~(\ref{LCpDeltaxi}) imply that to leading 
order in $\delta t$ and $\delta \theta^i$, the homogeneous and inhomogeneous
parts are given by 
\begin{equation}
\Lambda_{ti} = \ddot h_{ij}^{\mathrm{TT}}\delta\theta^j \delta t \, , \qquad
\Delta\xi_t = -\frac r2 \ddot h_{ij}^{\mathrm{TT}}\delta\theta^i\delta\theta^j 
\delta t \, , \qquad
\Delta\xi_i = -\frac 12 \ddot h_{ij}^{\mathrm{TT}}\delta\theta^j (\delta t)^2 
\, .
\end{equation}
Note that the homogeneous solution goes to a constant at large $r$ and that
part of the inhomogeneous solution scales as $r$. 
A completely analogous result for the homogeneous solution was derived in 
\cite{Helfer2014} using the Bondi framework near null infinity and the 
Newman-Penrose formalism.
This result implies that the holonomy of the affine transport may not be 
well suited for investigating angular momentum ambiguities around these
types of curves near null infinity, when the spacetime is dynamical.
The linear holonomy associated with parallel transport might be better
suited for these situations.

A different scaling for the homogeneous and inhomogeneous solutions was found
in \cite{FlanaganNichols2014} for spacetimes that undergo 
stationary-to-stationary transitions and for a class of 
observers following geodesic worldlines over long times separated by large
(or small) angles.
In this context, the holonomy scales as $1/r$, and the inhomogeneous solution 
approaches a constant near future null infinity for the curves considered.
For these larger curves, the higher-order terms in the expansion become 
relevant; thus, computing the holonomy by directly integrating the equations
of affine transport~(\ref{eq:AffineTrans}) becomes a more efficient and 
accurate method for computing the holonomy.  
It is a noteworthy property of stationary-to-stationary spacetimes
that holonomies around these large curves have can have a different 
scaling with $r$ than infinitesimal holonomies within the same loop do.
The requirement of stationarity at early and late times causes cancellations
between the holonomies in small regions such that the net result falls off
more rapidly with $r$ than the holonomies around the small area elements do.

\section{Conclusions and Discussion}
\label{sec:Conclusion}

In this paper, we used covariant bitensor methods to derive a series solution
for the generalized holonomy (both the homogeneous and inhomogeneous parts) 
around a geodesic triangle to any order, in terms of coincidence limits of
derivatives of the parallel propagator and Synge's world function.
We presented explicit results through fourth order in the vectors that
define the geodesic triangle.
Through third order in these vectors, the generalized holonomy (minus the 
identity map) around a parallelogramoid is just the sum of the generalized 
holonomies (again minus the identity maps) around the two triangles above 
and below its diagonal.
The lowest-order part of the linear holonomy around the parallelogramoid 
reproduces the standard textbook treatments.
The inhomogeneous part of the generalized holonomy is a higher-order quantity
for a connection without torsion.
We also computed higher-order corrections to both of these quantities, which
could be useful for estimating the errors in the leading-order expressions
for larger curves.

Because the inhomogeneous part of the generalized holonomy scales with area to 
the three-halves power times the Riemann tensor for a torsion-free connection, 
whereas the linear holonomy associated with parallel transport scales with the 
area times Riemann, the inhomogeneous solution will generally be less relevant 
for investigating the effects of spacetime curvature in infinitesimal regions 
than the linear holonomy is.
This scaling of the inhomogeneous solution with area seems relevant for
proposals to find spacetimes with ``torsion without torsion'' (see, e.g.,
\cite{Petti1986}).
In the language of this paper, these torsion-without-torsion solutions are 
spacetimes in which there is a nonvanishing inhomogeneous solution per unit 
area for a torsion-free connection.
Based on the results computed in this paper, as the unit area tends to zero, 
the inhomogeneous solution per unit area should vanish.
Thus, it seems that torsion without torsion would only be relevant for 
larger nonlocal spacetime regions.

\begin{acknowledgements}
We would like to thank \'Eanna Flanagan, Jordan Moxon, Leo Stein, and 
Peter Taylor for discussing aspects of this work with us; we thank Leo Stein in 
addition for providing comments on an earlier draft of the paper.
This work was supported by NSF Grants No.\ PHY-1404105 and PHY-1068541.
\end{acknowledgements}

\appendix

\section{The generalized holonomy for small geodesic triangles to fourth order 
in distance}\label{app:fourthorder}

The (symmetrized) coincidence limits needed to compute the holonomy of 
parallel transport around the triangle are given by
\bea
\Big[g^{a}{}_{b';(cdef)}\Big]&=&0 \, ,
\nnm\\\nnm
\Big[g^{a}{}_{b';(cde)f'}\Big]&=&-\frac{1}{4}R^a{}_{bf(c;de)}+\frac{1}{4}R^a{}_{bg(c}R^g{}_{de)f} \, ,
\\\nnm
\Big[g^{a}{}_{b';(cd)(e'f')}\Big]&=&\frac{1}{12}\Big(R^a{}_{b(c(e;f)d)}-R^a{}_{b(e(c;d)f)}\Big)+\frac{1}{2}R^a{}_{g(c(e}R_{f)d)b}{}^g \, ,
\\\nnm
&&+\frac{1}{4}\Big(R^a{}_{bg(c}R_{d)(ef)}{}^g-R^a{}_{bg(e}R_{f)(cd)}{}^g\Big) \, , 
\\\nnm
\Big[g^{a}{}_{b';c(d'e'f')}\Big]&=&\frac{1}{4}R^a{}_{bc(d;ef)}-\frac{1}{4}R^a{}_{bg(d}R^g{}_{ef)c} \, ,
\\\nnm
\Big[g^{a}{}_{b';(c'd'e'f')}\Big]&=&0 \, .
\eea
These [and some of the coincidence limits above (\ref{trihores})] have been 
obtained by differentiating the coincidence expansions presented in 
Refs.~\cite{OW,GDE} and applying Synge's rule \cite{Synge,PoissonReview}, 
while also employing the Bianchi identities and commuting derivatives of the 
Riemann tensor (to simplify the resulting expressions).
Following the calculation in Sec.~\ref{sec:GenHolHom}, the holonomy through 
fourth order is 
\bea
\Lambda^a{}_b&=&\delta^a{}_b+\frac{1}{2}R^a{}_{bvu}+\frac{1}{6}R^a{}_{bvu;(v+u)}+\frac{1}{8}R^a{}_{cvu}R^c{}_{bvu}
\\\nnm
&&+\frac{1}{48}\left\{\Big(R^a{}_{bvu;(2v+u)v}+R^a{}_{bcv}R^c{}_{(2v-3u)vu}\Big)-\Big(v\leftrightarrow u\Big)\right\}+O(5) \,,
\eea
where $u$ and $v$ appearing in index slots denote contractions of the Riemann 
tensor or its derivatives with $u^a$ and $v^a$.

To compute the inhomogeneous part of the generalized holonomy at fourth 
order, we employ the following (symmetrized) coincidence limits \cite{Synge}:
$$
\Big[\sigma^{;a}{}_{(bcde)}\Big]=0=\Big[\sigma^{;a}{}_{(b'c'd'e')}\Big]\,,\qq 
\Big[\sigma^{;a}{}_{(bc)(d'e')}\Big]=\frac{1}{6}R^a{}_{(de)(b;c)}+\frac{1}{2}R^a{}_{(de)(b;c)}
$$
$$
\Big[\sigma^{;a}{}_{(bcd)e'}\Big]=-\frac{1}{2}R^a{}_{(b|e|c;d)}
\,,\qq
\Big[\sigma^{;a}{}_{b(c'd'e')}\Big]=-\frac{1}{2}R^a{}_{(c|b|d;e)}\,.
$$
Using the results of Sec.~\ref{sec:GenHolHom}, the inhomogenous part has the
form 
\begin{equation}
\Delta\xi^a(\triangle_{u,v})=\left\{\left(\frac{1}{6}R^a{}_{vvu}+\frac{1}{12}R^a{}_{vuv;v}+\frac{1}{24}R^a{}_{uvu;v}\right)-\Big(v\leftrightarrow u\Big)\right\}+O(5)\,.
\end{equation}

\ifx\nobibtexrefs\undefined
\bibliographystyle{spphys}
\bibliography{Refs}
\else

\fi
\end{document}